# Robust Hybrid Image Watermarking based on Discrete Wavelet and Shearlet Transforms

## Abstract

With the growth of digital networks such as the Internet, digital media have been explosively developed in e-commerce and online services. This causes problems such as illegal copy and fake ownership. Watermarking is proposed as one of the solutions to such cases. Among different watermarking techniques, the wavelet transform has been used more because of its good ability in modeling the human visual system. Recently, Shearlet transform as an extension of Wavelet transform which is based on multi-resolution and multi-directional analysis is introduced. The most important feature of this transform is the appropriate representation of image edges. In this paper a hybrid scheme using Discrete Wavelet Transform (DWT) and Discrete Shearlet Transform (DST) is presented. In this way, the host image is decomposed using DWT, and then its low frequency sub-band is decomposed by DST. After that, the bidiagonal singular value decomposition (BSVD) is applied on the selected sub-band from Shearlet transform and the gray-scale watermark image is embedded into its bidiagonal singular values. The proposed method is examined on the images with different textures and resistance is evaluated against various attacks like image processing and geometric attacks. The results show good transparency and high robustness in proposed method.

Malihe Mardanpour [1] , Mohammad Ali Zare Chahooki [2]

[1,2] Electrical and Computer Engineering Department, Yazd University, Yazd, Iran. PO Box: 89195-741

[1] malihe.mardanpour@stu.yazd.ac.ir
[2] chahooki@yazd.ac.ir

## 1 Introduction

Rapid development of the Internet and multimedia technologies and also availability of editing and processing tools leads to easily distribution and manipulation of digital images without significant loss of quality. In such cases the security and protection of data has become an important issue [1]. Digital watermarking is one of the approaches for information hiding that is used for purposes such as identification and ownership protection of a digital product which in modern digital world it is called copyright protection. Procedure of image watermarking is such that secret information is embedded in the host image and then in receiver side this image, which contains secret message, is received for owner identification or verification. The main challenge in this area is to achieve the least distortion in image quality and protection of embedded data against intentional removal attacks. So far, none of the methods have been able to cover all the aspects of watermarking but they try to achieve balanced optimal values for robustness, transparency and capacity, because these three attributes are mutually orthogonal to each other [2].

Watermarking techniques based on the information needed during extraction of watermark divided into two categories: blind, non-blind. Methods which need to host image in extraction process are non-blind and those which does not need are blind methods [3]. Other classification for watermarking methods is based on the domain that watermark is inserted in, which categorized into two spatial [4] and the frequency [5] domains. In the spatial class, watermark is directly embedded in host image components. These methods have low computational cost and simple implementation, but are fragile against attacks.In frequency domain algorithms watermark is embedded by changing the coefficients magnitude of image in the frequency domain. These methods can embed more information and also have more resistant against malicious manipulation and image processing attacks [6]. Some of the well-known frequency domain transforms are Discrete Fourier Transform (DFT), Discrete Cosine Transform (DCT), and DWT [7]. DCT-based methods are robust against simple image processing attacks like low-pass filtering and blurring but they can not resistant against attacks such as rotation, resizing, and cropping. Watermarking techniques based on DFT in addition to image processing attacks are robust against some geometric attacks.

DWT compared to other frequency transforms is more efficient because it can be expressed in time and frequency and its multi-resolution representation [8]. Having features such as similarity of DWT to the human visual system (HVS) and good modeling of HVS, space frequency localization and multi-resolution representation cause DWT is used in watermarking and image processing researches [9], as well as it is suitable for identifying frequency regions of the image signal such that the watermark can be effectively embedded [10]. Indeed advantages of DWT, it has disadvantages like shift invariant, which occurs because of using down-samplers after each filtering step. This leads significant changes with small changes in wavelet coefficients. This makes inaccurate extraction of the watermark. Redundant Discrete Wavelet Transform (RDWT) is designed for this problem [11]. In [11], makbol et.al. proposed a method based on RDWT-SVD which the gray-scale watermark is embedded in singular values of sub-bands from RDWT decomposition. RDWT with other transforms such as Lifting Wavelet Transform (LWT) [12], Integer Wavelet Transform (IWT) [13], Wavelet Packet Transform (WPT) [14] and Complex Wavelet Transform [15] are extensions of wavelet transforms.

The main limitation of the Wavelet transform is its low ability to capture the directional information. To overcome this, multi-scale and directional representations are used to capture the intrinsic geometrical structures such as smooth contours in natural images. These multi-scale and directional representations perform scale and directional analysis using geometric transform and local directional, respectively.

In this context, some of these multi-scale transforms are Ridgelet [16], Contourlet [17], Curvelet [18], and Shearlet [19]. These transforms are used in various image processing applications like denoising [18,20,21,22], compression [23,24], segmentation [25, 26], image retrieval [27,28,29 ] and feature extraction [30,31,32,33]. Although most of the multi-scale representations are overcomplete and have high computational complexity, but they are used since they can represent fine details of the natural images [34].

Since Curvelet transform is not built directly in discrete domain, it can not provide a multi-resolution representation of the geometry. Contourlet transform has less clear directional features rather than Curvelet transform. In recent years, DST was proposed to improve the existing methods. DST provides a new representation for discrete multi-scale directional [35]. Shearlet compared to transforms like Contourlet, Curvelet, and Ridgelet obtains a unique combination of mathematical ridgeness and computational performance which in dealing with edges is efficient and computationally effective [36]. Indeed, Wavelets have wide impact in image processing, but they can not represent optimally objects with anisotropic elements such as lines or curved structure. The reason of it is that Wavelets are non-geometrical and do not use regularity curved edges [37]. Ahmaderaghi et.al. in [35] proposed a watermarking scheme based on spread spectrum using Shearlet. They also in [38] presented a watermarking method based on chou's model using multi-resolution and multi-directional features of Shearlet transform. In [39], binary image divided into non overlapping blocks and Ridgelet transform is applied on each block, then watermark is embedded in the coefficients matrix of Ridgelet with high variance. A robust watermarking technique is introduced in ridgelet domain in [40]. In this approach watermark is inserted into the selected blocks of the host image by modulating the ridgelet coefficients which shows the most energetic direction. In [41] a digital image watermarking method presented based on Ridgelet transform which in extraction process Independent Component Analysis (ICA) is exploited. An adaptive watermarking algorithm is introduced with Curvelet in [42]. First, the host image is transformed into curvelet domain and then watermark is embedded in intermediate frequency sub-bands coefficients of Curvelet. In [43], an algorithm based on Curvelet is provided which extracts position for insertion of watermark according to characteristics of the HVS. A non-blind watermarking technique using Contourlet transform is proposed in [44] that is based on the similarity approach. In this way, sub-bands from Contourlet decomposition is divided into $8\times8$ blocks and similarity coefficients between the watermark blocks and sub-bands are calculated and watermark is embedded in the block of directional sub-bands with the highest similarity. In [45-47], watermarking methods using Contourlet and SVD are presented. In [48], a hybrid watermarking using Wavelet transform and Contourlet is proposed. In this method, host image is decomposed by Contourlet and then three-level Wavelet transform is applied on the low-frequency sub-bands from Contourlet. Finally watermark is inserted in the low-frequency sub-band resulted from Wavelet decomposition.

Factorization of a matrix in Linear Algebra is a decomposition of a matrix into a product of a matrices. SVD is one of the methods of matrix factorization which is used for signal spectral decomposition. This transform is widely used in image processing applications such as image hiding [49], image compression [50], noise reduction [51], and image watermarking [6,7,11,13] . The important feature of SVD is good stability of singular values of images, such that small variations does not make significant changes in singular values of images [6]. For the first time in 2002, Liu et.al. proposed a SVD-based watermarking [52]. BSVD can be viewed as a development for SVD which is employed for matrix decomposition. Singular values obtained from both methods are same, but the procedure of calculation for singular values in BSVD is more efficient than SVD [53].

In this paper, a watermarking method is proposed based on DWT and DST with BSVD. The main innovation in this study is hierarchical using of DWT and DST.

In this approach, watermark is embedded in bidiagonal singular values of the sub-band result from consecutive applying DWT and DST. So that, host image is decomposed by DWT, first. Then DST is applied on the low-frequency sub-band of it. Among sub-bands result from DST decomposition, one of them is selected and watermark is inserted directly in bidiagonal singular values of it. The proposed method is tested on various images with different texture characteristics. Also a wide range of attacks with different parameters are applied on watermarked images. The results of the experiments show high imperceptibility of the proposed method, as well as its robustness against attacks. This approach can be used for applications having images with various texture properties.

In the following of the paper in Section 2, an overview of definition and basic concept of DWT, DST and BSVD is provided. Our DWT-DST method will be discussed in Section 3. In Section 4, experimental results will be described and in the last section, conclusions and future studies will be explained.

## 2 Background Review

In this section, some fundamental concepts of the transforms used in this study is explained. In the following sub-sections DWT, DST, and BSVD are described respectively.

### 2.1 DWT

DWT is popularly used in signal processing applications. The idea of DWT is multi-resolution analysis which decomposes an image into frequency channels with fixed bandwidth on a logarithmic scale. Similarity to a structure according to a certain resolution and decomposition at each level is one of the advantages of DWT. It can be implemented as a multi-step transform. An image is decomposed into four sub-bands, LL, LH, HL, HH by 1-level DWT. The first letter of each sub-band corresponds to performing a frequency low-pass or high-pass operation on the rows and second letter allude to the filter which is applied on the columns. LH, HL and HH show finest scale wavelet coefficients and LL contains coarse-level ones. LL sub-band can be decomposed in next level of decompositions to obtain another level of decomposition. Analysis process on the LL sub-band continues until reach to the desired number of decomposition level for a specific purpose [54]. Applying 1-level DWT on a given image f(x, y) is written as in equation (1-4):

$$\mathrm{LL} = [(\mathrm{f}(\mathrm{x},\mathrm{y}) \times \Phi - \mathrm{x}\Phi - \mathrm{y})(2\mathrm{n}, 2\mathrm{m})]_{(\mathrm{n},\mathrm{m})\in \mathrm{Z}^2} \quad (1)$$

$$LH = [(f(x, y) \times \Phi - x\psi - y)(2n, 2m)]_{(n,m)\in Z^2} \quad (2)$$

$$HL = [(f(x, y) \times \psi - x\Phi - y)(2n, 2m)]_{(n,m)\in Z^2} \quad (3)$$

$$HH = [(f(x, y) \times \psi - x\psi - y)(2n, 2m)]_{(n,m)\in Z^2} \quad (4)$$

In above equations Φ(t) and ψ(t) are scaling and wavelet functions respectively. In figure 1, decomposition of 1-level DWT and the four sub-bands is shown.

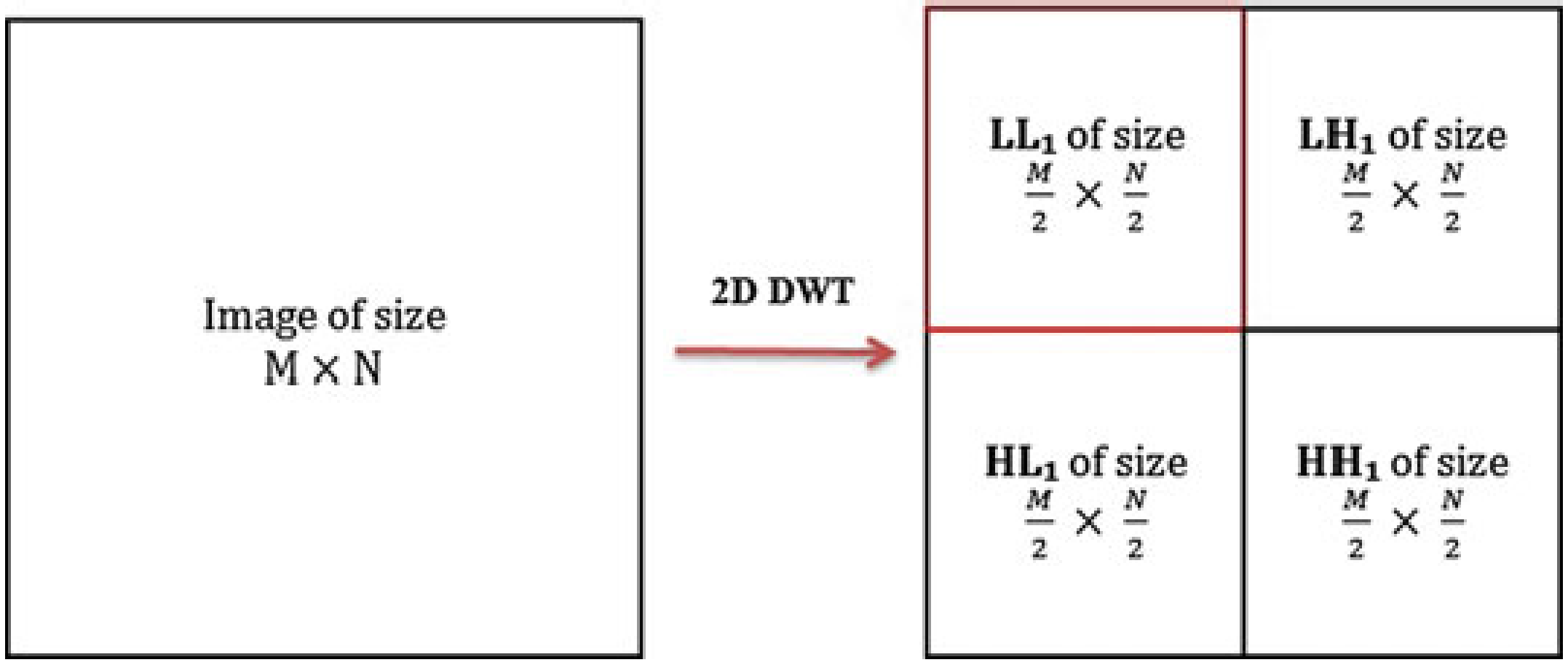

Figure 1. 1-level DWT decomposition on a given image of size M×N [55].

## 2.2 DST

In the last decade, a new generation of multi-scale transforms emerged which combines classical multi-resolution analysis power with the high performance directional information processing ability. Shearlet, Curvelet, and Contourlet are examples of such transforms. Unlike classic Wavelets the elements of these transforms create a pyramid of well-localized waveforms such that contains different orientations with high anisotropic shapes in addition to various scales and locations. According to their rich structure they can overcome to the poor directional of the multi-scale systems and can be used in the latest algorithms of the signal and image processing. Shearlet presents unique combination of significant features: having understandable and simple mathematical structure which is taken from the affine systems theory. It provides optimally sparse representation, also, its directionality is controlled by shear matrices instead of rotations. Shearlet transform is employed in many problems such as applied mathematics, signal processing like operators decomposition, inverse problems, edge detection, and image restoration [56]. Shearlet is a framework of the affine system which extractsgeometrical features of multi-dimensional signals [57]. This transform is an affine system and includes a shearlet function which is parametrized with scaling, shear and translation such that shear parameter captures direction of the singularities [35].

For image I, Shearlet transform is a mapping in the form of relation (5):

$$\mathrm{I} \rightarrow \mathrm{SH}_{\psi}\mathrm{I}(\mathrm{a}, \mathrm{s}, \mathrm{x}) \qquad (5)$$

Which depends to scale $a$>0, direction $s$ and location $x$. Shearlet transform is expressed as equation (6):

$$\mathrm{SH}_{\psi}\mathrm{I}(\mathrm{a}, \mathrm{s}, \mathrm{x}) = \int \mathrm{I}(\mathrm{y})\Psi_{\mathrm{as}}(\mathrm{x}-\mathrm{y})\mathrm{dy} = \mathrm{I} \times \Psi_{\mathrm{as}}(\mathrm{x}) \qquad (6)$$

Shearlets are constructed by dilating, shearing and translation which each mother function $\psi \in L_2(R^2)$ is obtained by equation (7).

$$\Psi_{j,k,m}(x) = |\det A|^{\frac{j}{2}}\Psi(S^k A^j x - 1) \qquad (7)$$

$A$ and $S$ are 2×2 invertible matrices that represent geometrical transformations and dilation as relation (8).

$$\mathrm{A} = \begin{pmatrix} \mathrm{a} & 0 \\ 0 & \sqrt{\mathrm{a}} \end{pmatrix}, \mathrm{S} = \begin{pmatrix} 1 & \mathrm{s} \\ 0 & 1 \end{pmatrix} \qquad (8)$$

So that, for each mother function we have DST as equation (9):

$$SH\left\{\Psi_{j,k,m} = 2^{\frac{3}{4}j}\Psi(s_k A_2 j - m) : j, k \in \mathbb{Z}^2\right\} \qquad (9)$$

According to the good capability of DST for capturing directional characteristics and good localization, DST is a suitable choice for watermarking [35]. In figure 2, tiling of the frequency plane for DST is illustrated.

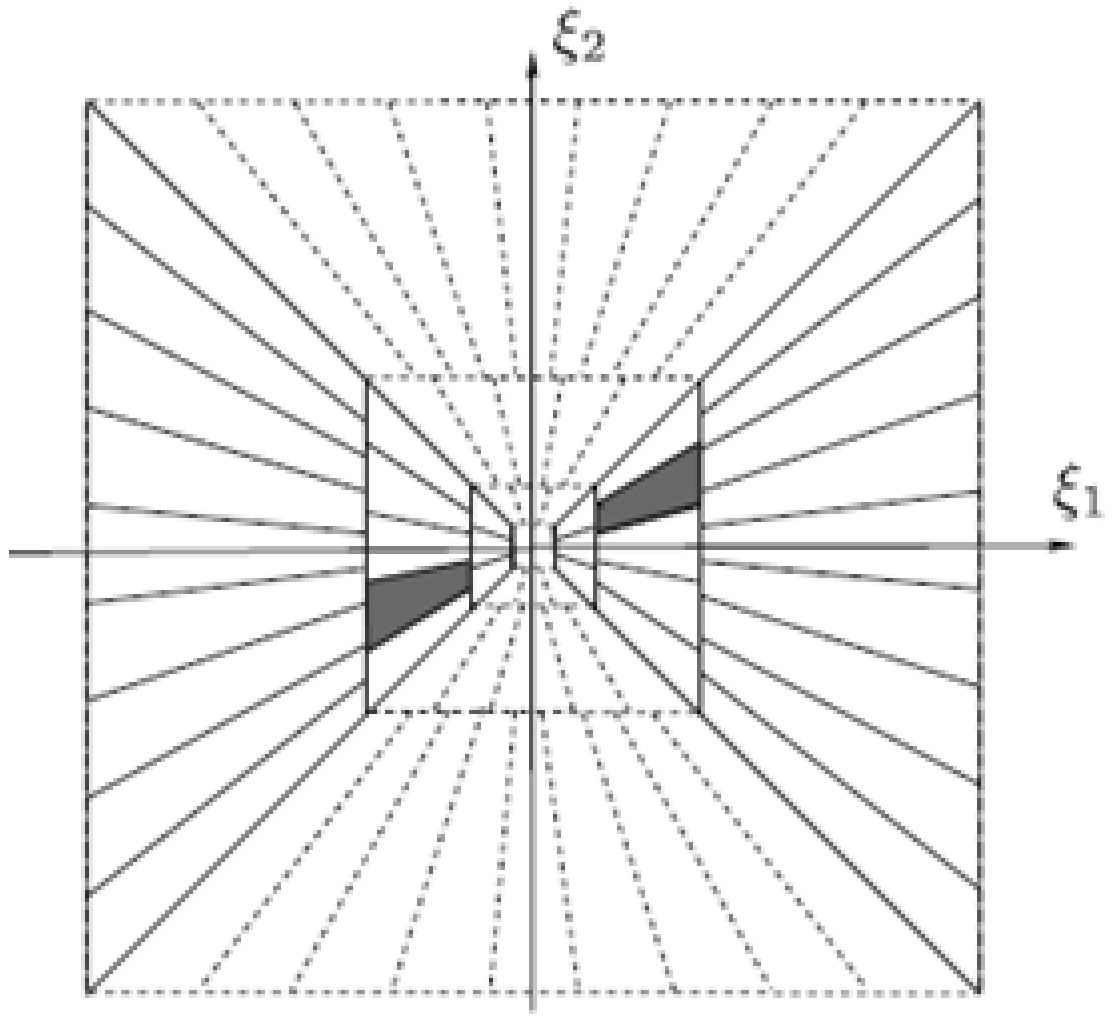

Figure 1. The tiling of the frequency plane $\mathbb{R}^2$ induced by the shearlets [56].

Figure 3, shows the filter bank decomposition for Shearlet transform. According to that, first the image is decomposed to a low-pass sub-band and a band-pass sub-band by Laplacian Pyramid (LP). Then, band-pass sub-band which demonstrates the difference between the original image and the sub-band low-pass filters is delivered to an appropriate shearing filter to complete multi-directional decomposition. This process is performed continuously on the low-pass sub-bands to obtain a multi-scale decomposition [58].

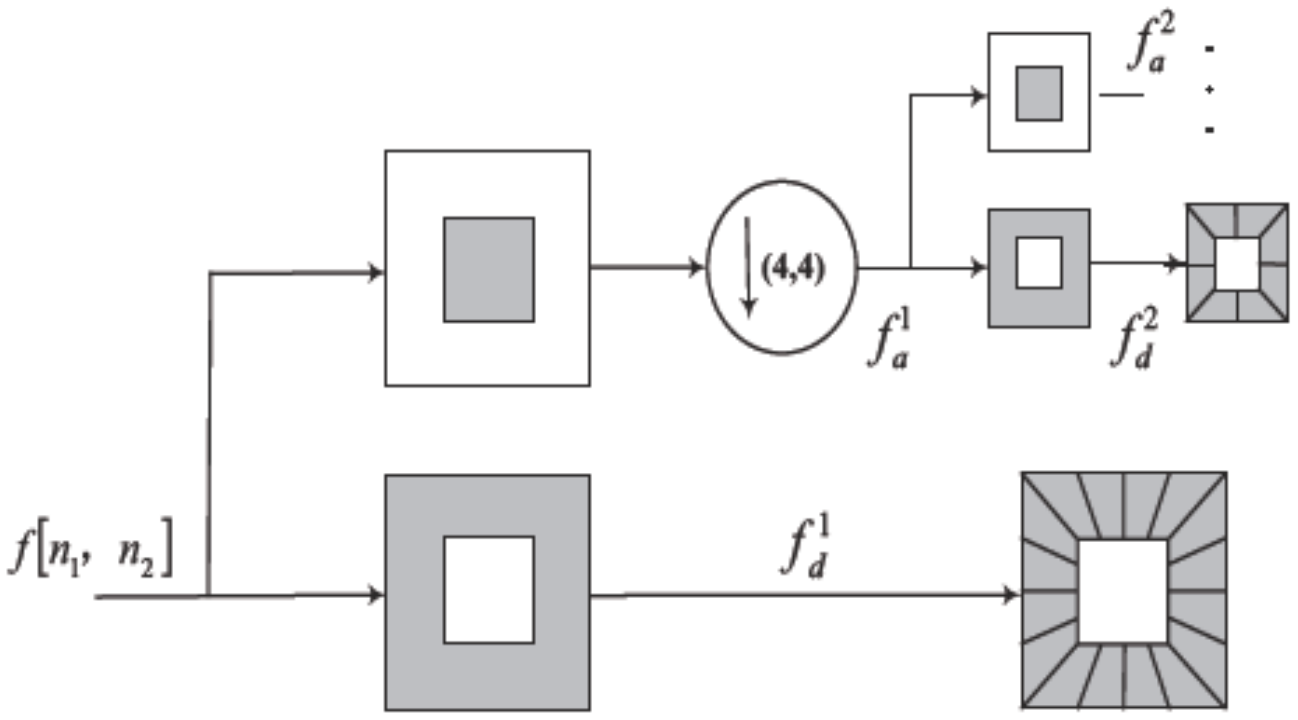

Figure 2. The filter bank decomposition of Shearlet transform [58].

DST decomposes an image into $2\times2^{nj}+1$ directional sub-bands in horizontal cone and vertical cone in each scale respectively which $j$ is decomposition level and $n_j$ is direction parameter (*ndir*). For example when *ndir*=[2 1 0], total number of directions in both directions, horizontal cone and vertical cone, is 2×(9+5+3)=34 [57].

## 2.3 BSVD

SVD is a matrix factorization method which is used in different image processing applications such as image hiding, image compression, noise reduction and also image watermarking. Several image watermarking techniques have been presented using this decomposition lonely and also with other transforms. Since performing SVD on images needs a lot of computation, so it is used in hybrid schemes. The SVD of an image A with size $m\times m$ is in the form of $A = USV^T$, where $U$ and $V$ are orthogonal matrices, and $S = \text{diag}(\lambda_i)$ is a diagonal matrix of singular values $\lambda_i$, $i = 1, . . . , m$, which are in decreasing order.

From features of SVD which cause to use it in digital watermarking methods can mention as (1) When image is distorted a little, many of its singular values does not change, and (2) Singular values depict natural algebraic image characteristic [54].

BSVD is considered a spectral decomposition which is based on SVD. Actually these two decompositions supply similar singular values. Although, the procedure of computation for singular values in BSVD and SVD are different [53]. BSVD calculation of a given matrix, A, is as below:

1) bidiagonalization of A, which is performed by equation (10):

$$A=U_A \times B \times V_A^T \qquad (10)$$

where $U_A$ is an $m\times n$ orthonormal matrix, $V_A$ is an $n\times n$ unitary matrix, and $B$ strictly an upper bidiagonal matrix of size $n\times n$.

2) applying SVD on B, which is as equation (11):

$$B = U_B S V^T_B \qquad (11)$$

where $U_B$ and $V_B$ are unitary matrix and $S$ is defined as equation 12:

$$S= \text{diag}(\sigma_1, \sigma_2, . . . , \sigma_r) \qquad (12)$$

where $\sigma_i$, are the singular values of the matrix $B$ with $r = min(m, n)$ and satisfying $\sigma_1 \geq \sigma_2 \geq \cdots \geq \sigma_r$.

So, BSVD decomposition of matrix A is obtained by replacing equation (11) into (10). So, it can be written as equation (13):

$$A = U_A U_B S V^T_B V^T_A \qquad (13)$$

The difference between SVD and BSVD is the routine of singular values calculation, which BSVD is better than SVD in terms of performance. Number of keys is another advantage of BSVD over SVD, such that BSVD gives four keys while SVD have two keys [53].

## 3 The proposed DWT-DST watermarking

In this paper, DWT and DST are exploited with BSVD spectral decomposition for improving robustness and transparency. Host image is decomposed by 1-level DWT into four sub-bands LL,LH,HL and HH. Then DST is applied on low-frequency sub-band LL. Watermark is embedded directly in bidiagonal singular values resulted from BSVD on the selected sub-band which is outcome of DST decomposition.

### 3.1 Watermark embedding algorithm

Steps of embedding watermark into host image are as follows:

1- Host image, A, is first decomposed by 1-level DWT. So that, four sub-bands named LL,LH,HL, and HH are achieved.

2- Then, DST is applied on the low-frequency sub-band LL. Number of scales in DST is set to 3 and vector of Shearlet levels is determined to [0,1,1]. Therefore, 21 sub-bands are totally obtained. So, in first scale, four sub-bands, and eight sub-bands in second and third scales are achieved.

3- The sub-band in the first scale and vertical orientation is selected as embedding host frequency. This sub-band is selected based on trial and error approach. Then BSVD is applied on this sub-band according to equation (14).

$$A = U_A U_B S V^T{}_B V^T{}_A \qquad (14)$$

So, five matrices $U_A$, $U_B$, $S$, $V_A$ and $V_B$ are obtained. $S$ is the bidiagonal singular values of the image.

4- Watermark is embedded in bidiagonal singular values of $S$ and then SVD is performed according to equation (15):

$$S + \alpha W = U_w S_w V_w^T \qquad (15)$$

Where $\alpha$ stands for the scaling factor which is balancing factor between transparency and robustness and is adjusted by it.

5- The new modified DST coefficients are evaluated as equation (16):

$$A^{new} = U_A U_B S_w V_B^T V_A^T \qquad (16)$$

6- Finally, inverse DST and inverse DWT is performed on $A^{new}$ respectively. So, watermarked image, $A_w^*$, is achieved.

### 3.2 Watermark extraction algorithm

Extracting watermark from watermarked image is done as below:

1- Watermarked image is decomposed by 1-level DWT and four sub-bands LL, LH, HL, and HH are achieved. Then, DST is applied on LL sub-band with the same characteristics as mentioned in the first step of embedding algorithm.

2- BSVD is applied on the sub-band in the first scale and in the vertical orientation.

$$A_W^* = U_A^* U_B^* S^* V_B^{*T} V_A^{*T} \quad (17)$$

3- $D^*$ is calculated as equation (18):

$$D^* = U_W S^* V_W^T \quad (18)$$

4- Extracted watermark ,$W^*$, is computed by reverse embedded formula as equation (19):

$$W^* = (D^* - S)/\alpha \quad (19)$$

## 4 Experimental result

The proposed DWT-DST scheme is implemented using Matlab. ShearLab 3D [59] is used in the proposed method for DST calculations. Gray-scale images airplane (F-16), lenna and baboon from classic set of Stirmark benchmark and other images: peppers, barbara, boat, couple , elaine, man, sailboat, and lake which have different texture properties are used as host images. Size of all these images is 512×512 pixels. In our experiments, the watermark is a gray-scale image, “copyright logo”, with size of 256×256 pixels. Table 1 shows these images respectively.

To show effectiveness of our proposed scheme, transparency and robustness against different attacks which are the most important features of any watermarking system is quantified through Peak Signal-to-Noise Ratio (PSNR) and Normalized cross Correlation (NC) respectively. PSNR shows degree of visual similarity between watermarked and host image. It is computed as equation (20).

$$PSNR = 10\ log_{10}\left[\frac{max(x(i,j))^2}{MSE}\right] \quad (20)$$

Which Mean Square Error (MSE) is achieved from equation (21):

$$MSE = \frac{1}{m*n}\sum_{i=1}^{m}\sum_{j=1}^{n}[x(i,j) - y(i,j)]^2 \qquad (21)$$

Recently, another metric for similarity comparison between two images is introduced. It is called Structural Similarity (SSIM) and is designed to improve traditional methods like PSNR, which has been proved to be inconsistent with human eye perception. SSIM values are in range of [-1,1], where 1 is acquired when two image is identical. SSIM is calculated on different windows of an image. SSIM is calculated as equation (22):

$$SIM(x,y) = \frac{(2\mu_x\mu_y+c_1)(2\sigma_{xy}+c_2)}{(\mu_x^2+\mu_y^2+c_1)(\sigma_x^2+\sigma_y^2+c_2)} \qquad (22)$$

where $\mu_x$ and $\mu_y$ are the average of $x$ and $y$, and $\sigma_x$ and $\sigma_y$ are variance of $x$ and $y$. $\sigma_{xy}$ is the covariance value between $x$ and $y$. $c1$and $c2$ are two variables to stabilize the division with weak denominator [38].

Table 1 illustrates host images and watermark image. Bottom of all host images is written PSNR and SSIM values of corresponding watermarked images which is provided by proposed method. As we can see, all values for imperceptibility requirement are high, thus it could be said that proposed method is transparent.

**Table1. Host images with PSNR and SSIM of their corresponding watermarked image and watermark image.**

| | | | |
|---|---|---|---|
| 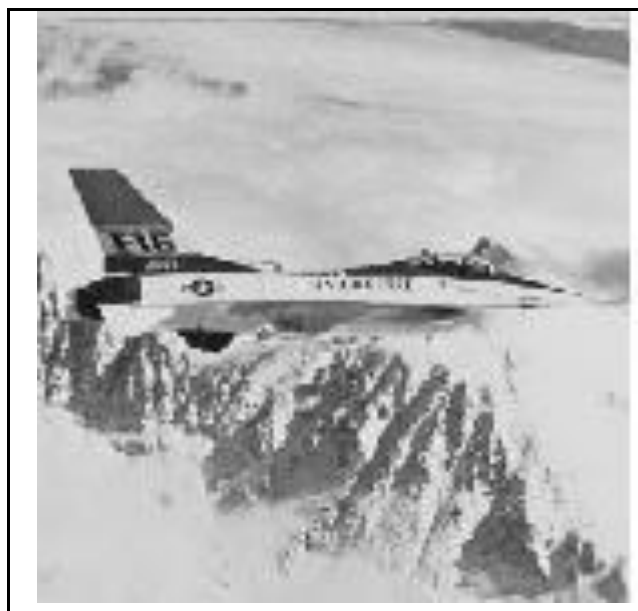 PSNR =69.14 db SSIM=1 | 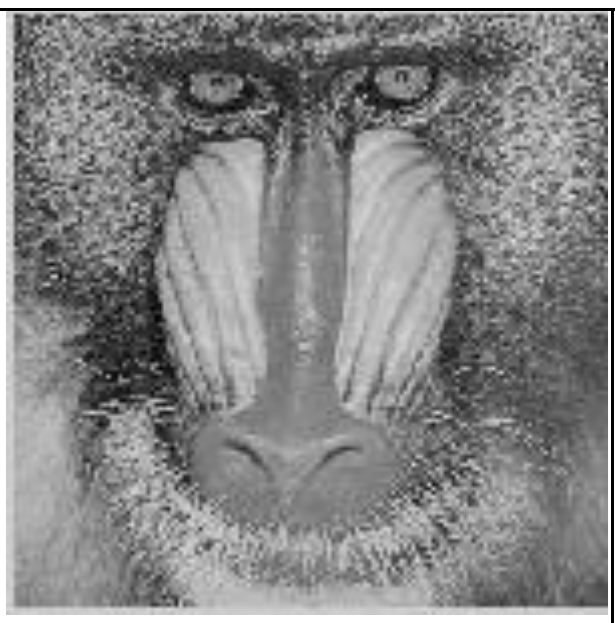 PSNR =73.79 db SSIM=1 | 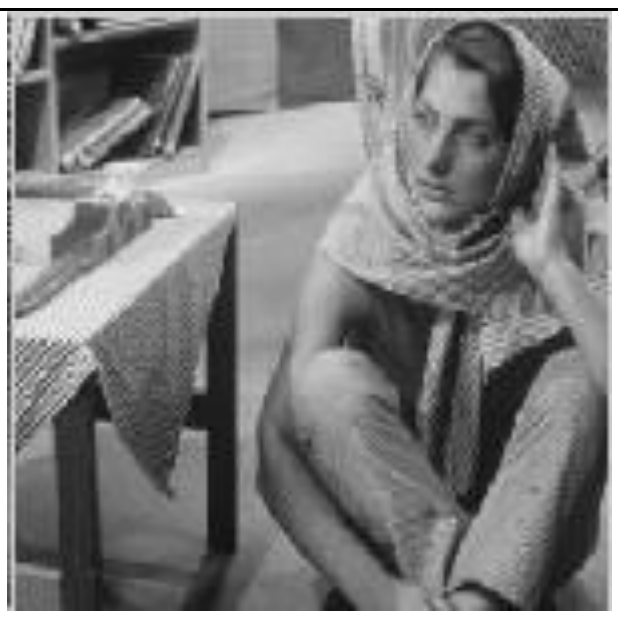 PSNR =68.27 db SSIM=1 | 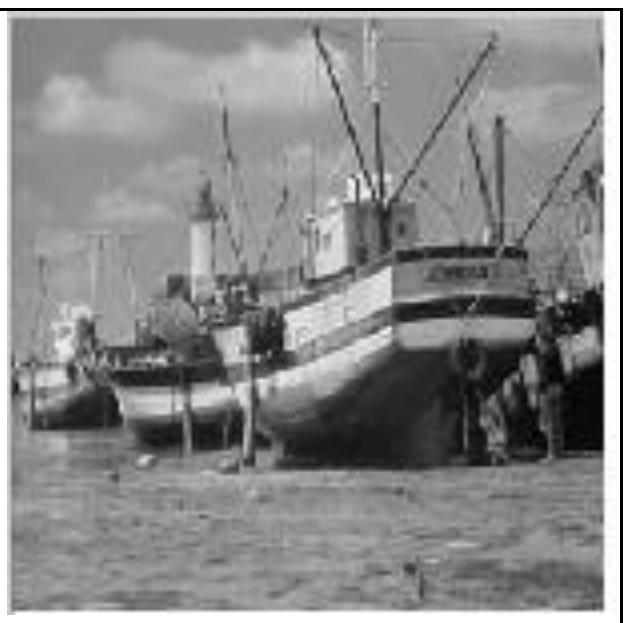 PSNR =69.91 db SSIM=1 |
| 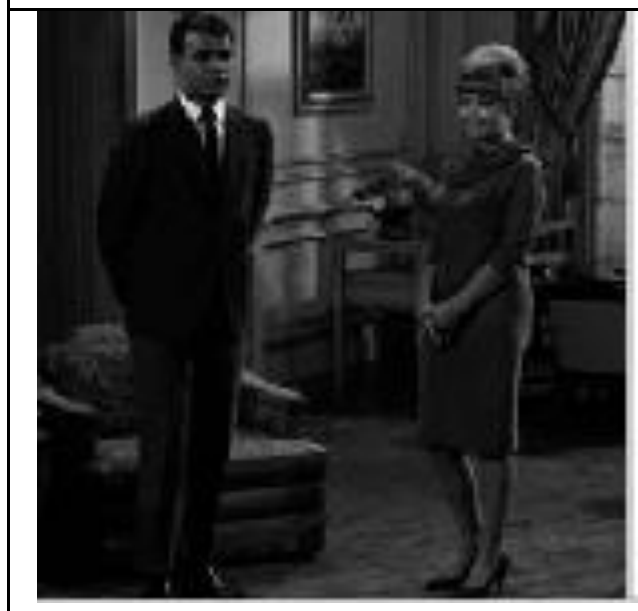 PSNR =65.65 db SSIM=0.999 | 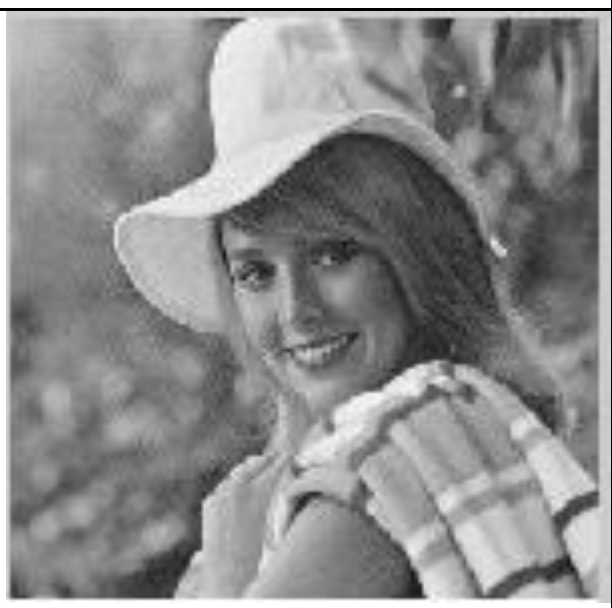 PSNR =67.88 db SSIM=1 | 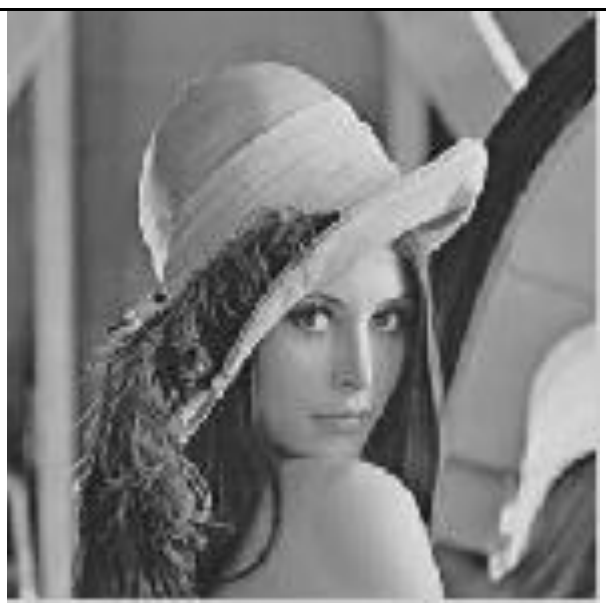 PSNR =67.54 db SSIM=1 | 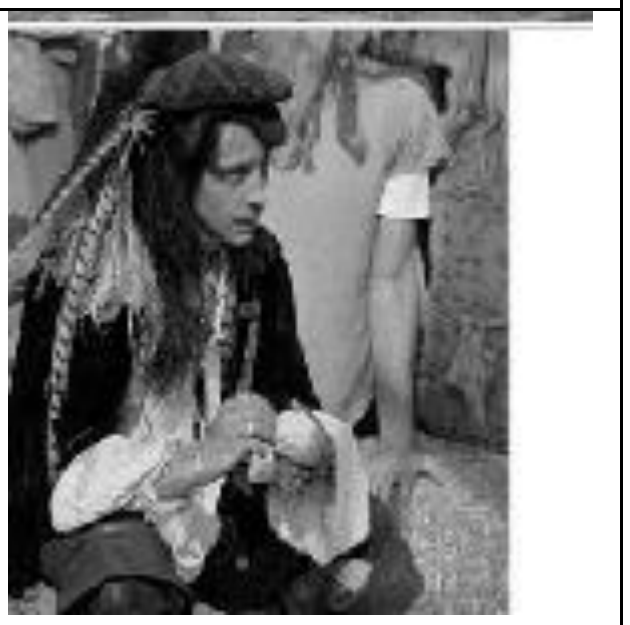 PSNR =73.32 db SSIM=1 |
| 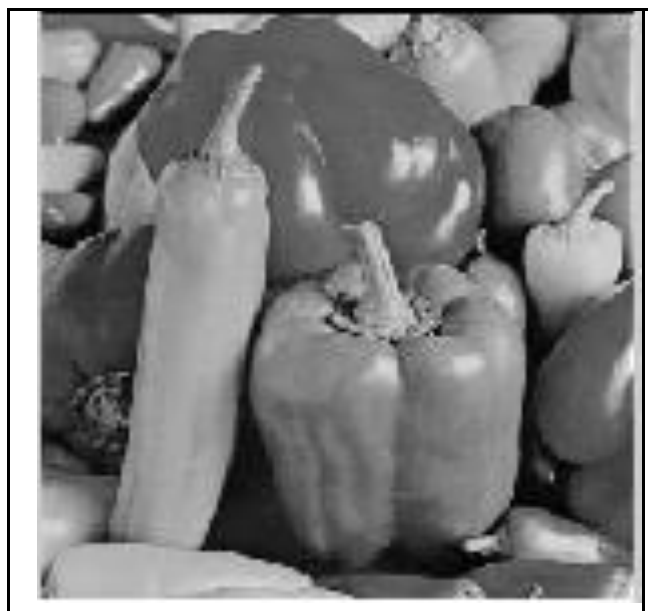 | 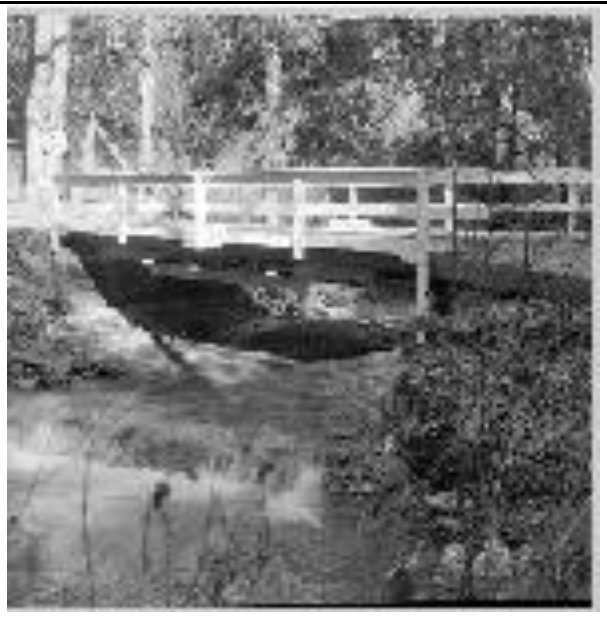 | 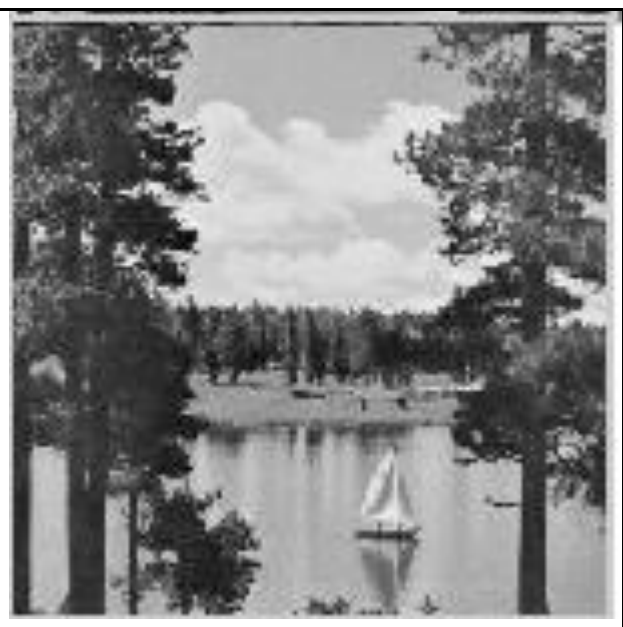 | 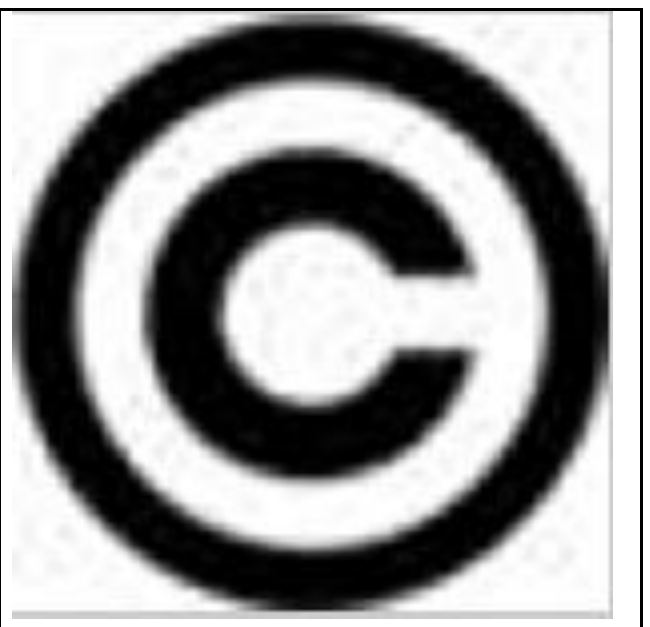 |
| PSNR =68.25 db<br>SSIM=1 | PSNR =74.62 db<br>SSIM=1 | PSNR =74.8 db<br>SSIM=1 | watermark logo |

As said before, robustness is another important requirement of watermarking method which is measured by NC. NC shows similarity between original watermark and extracted one from watermarked image attacked. It is calculated through equation (23):

$$NC(w,\overline{w}) = \frac{\sum_{i=1}^{m}\sum_{j=1}^{n}[w(i,j)-\mu_w][\overline{w}(i,j)-\mu_{\overline{w}}]}{\sqrt{\sum_{i=1}^{m}\sum_{j=1}^{n}[w(i,j)-\mu_w]^2}\sqrt{\sum_{i=1}^{m}\sum_{j=1}^{n}[\overline{w}(i,j)-\mu_{\overline{w}}]^2}} \quad (23)$$

In this study according to the results from testing the proposed method, the best value for scaling factor in embedding equation (15) which make a balance betwee PSNR and NC is 0.008.

Different attacks like image processing and geometric attacks have been performed on all of the watermarked images and then robustness of them has been evaluated using NC. Types of attacks utilized here include: Average Filter (AF), Gaussian low-pass Filter (GP), Median Filter (MF), Gaussian Noise (GN), Speckle Noise (SN), Salt & Pepper noise (SP), Blurring (BL), Gamma Correction (GC), Histogram Equalization (HE), Motion Blur (MB), Sharpening (SH), JPEG compression (JPEG), Crop (CR), Rotation (RO), Scaling (SC), Translation (TR), Shearing (SE), and Flip (FL).

## 4.1 Comprehensive attach evaluations

In table 2, NC values for all host images through each attack are depicted. According to values, the proposed method is robust against geometric and non-geometric attacks.

Table2. NC values of different attacks on all of the host images that watermarked by a 256×256 gray-scale image. For simplicity, decimal points have been eliminated in this table, to get the real numbers they should all be multiplied by $10^{-2}$.

| Attack | airplane | baboon | barbara | boat | couple | elaine | lenna | man | peppers | river | tree | Attack | airplane | baboon | barbara | boat | couple | elaine | lenna | man | peppers | river | tree |
|---|---|---|---|---|---|---|---|---|---|---|---|---|---|---|---|---|---|---|---|---|---|---|---|
| AF 3×3 | 98 | 98 | 98 | 99 | 99 | 99 | 99 | 98 | 99 | 98 | 98 | BL 0.2 | 99 | 99 | 99 | 99 | 99 | 99 | 99 | 99 | 99 | 99 | 99 |
| AF 5×5 | 97 | 97 | 98 | 98 | 98 | 98 | 98 | 97 | 98 | 97 | 97 | BL 0.5 | 99 | 99 | 99 | 99 | 99 | 99 | 99 | 99 | 99 | 99 | 99 |
| AF 10×10 | 85 | 84 | 91 | 90 | 95 | 91 | 90 | 87 | 89 | 88 | 86 | BL 0.8 | 99 | 99 | 99 | 99 | 99 | 99 | 99 | 99 | 99 | 99 | 99 |
| GP 3×3 | 99 | 99 | 99 | 99 | 99 | 99 | 99 | 99 | 99 | 99 | 99 | BL 1 | 99 | 98 | 99 | 99 | 99 | 99 | 99 | 99 | 99 | 99 | 99 |
| GP 5×5 | 99 | 99 | 99 | 99 | 99 | 99 | 99 | 99 | 99 | 99 | 99 | BL 1.1 | 99 | 98 | 99 | 99 | 99 | 99 | 99 | 99 | 99 | 99 | 99 |
| MF 2×2 | 99 | 99 | 99 | 99 | 99 | 99 | 99 | 99 | 99 | 99 | 99 | GC 0.8 | 22 | 68 | 73 | 60 | 97 | 19 | 85 | 94 | 79 | 82 | 75 |
| MF 3×3 | 99 | 98 | 99 | 99 | 99 | 99 | 99 | 99 | 99 | 98 | 99 | GC 0.9 | 85 | 98 | 99 | 98 | 99 | 98 | 98 | 98 | 99 | 99 | 98 |
| MF 4×4 | 98 | 98 | 98 | 99 | 99 | 99 | 99 | 98 | 99 | 98 | 98 | HE | 97 | 96 | 98 | 97 | 96 | 97 | 98 | 99 | 98 | 98 | 99 |
| MF 5×5 | 98 | 97 | 98 | 98 | 99 | 98 | 98 | 98 | 98 | 98 | 98 | MB(15,20) | 71 | 77 | 80 | 78 | 91 | 79 | 80 | 76 | 79 | 79 | 74 |
| MF 10×10 | 94 | 89 | 96 | 92 | 97 | 96 | 96 | 94 | 96 | 92 | 92 | MB(15,45) | 72 | 76 | 78 | 78 | 90 | 79 | 79 | 75 | 78 | 76 | 73 |
| CR 30% | 32 | 79 | 44 | 31 | 43 | 45 | 43 | 55 | 55 | 68 | 57 | SH 0.1 | 99 | 98 | 99 | 99 | 99 | 99 | 99 | 99 | 99 | 99 | 99 |
| CR 50% | 50 | 92 | 75 | 49 | 78 | 70 | 54 | 84 | 87 | 92 | 80 | SH 0.2 | 99 | 99 | 99 | 99 | 99 | 99 | 99 | 99 | 99 | 99 | 99 |
| CR 70% | 75 | 97 | 84 | 66 | 91 | 77 | 65 | 94 | 91 | 97 | 89 | SH 0.5 | 99 | 99 | 99 | 99 | 99 | 99 | 99 | 99 | 99 | 99 | 99 |
| CR 90% | 86 | 96 | 83 | 79 | 92 | 84 | 75 | 96 | 93 | 98 | 92 | SH 0.8 | 99 | 99 | 99 | 99 | 99 | 99 | 99 | 99 | 99 | 99 | 99 |
| RO 5° | 97 | 98 | 98 | 99 | 99 | 99 | 99 | 99 | 98 | 98 | 98 | SH 1 | 99 | 99 | 99 | 99 | 99 | 99 | 99 | 99 | 99 | 99 | 99 |
| RO 30° | 97 | 98 | 98 | 98 | 98 | 98 | 98 | 96 | 97 | 97 | 98 | JPEG Q=5 | 99 | 99 | 99 | 99 | 99 | 99 | 99 | 99 | 99 | 99 | 99 |
| RO 45° | 98 | 99 | 98 | 97 | 98 | 97 | 96 | 96 | 98 | 97 | 98 | JPEG Q=10 | 99 | 99 | 99 | 99 | 99 | 99 | 99 | 99 | 99 | 99 | 99 |
| RO 70° | 99 | 98 | 98 | 99 | 97 | 98 | 98 | 97 | 98 | 98 | 97 | JPEG Q=20 | 99 | 99 | 99 | 99 | 99 | 99 | 99 | 99 | 99 | 99 | 99 |
| RO 110° | 95 | 97 | 97 | 97 | 97 | 98 | 97 | 98 | 98 | 96 | 97 | JPEG Q=30 | 99 | 99 | 99 | 99 | 99 | 99 | 99 | 99 | 99 | 99 | 99 |
| RO -10° | 98 | 98 | 98 | 98 | 98 | 98 | 98 | 98 | 99 | 98 | 97 | JPEG Q=80 | 99 | 99 | 99 | 99 | 99 | 99 | 99 | 99 | 99 | 99 | 99 |
| RO -50° | 97 | 97 | 96 | 98 | 96 | 97 | 92 | 95 | 96 | 97 | 98 | JPEG Q=90 | 99 | 99 | 99 | 99 | 99 | 99 | 99 | 99 | 99 | 99 | 99 |
| RO -80° | 94 | 97 | 98 | 96 | 98 | 98 | 98 | 98 | 98 | 96 | 98 | GN(0,0.005) | 99 | 99 | 99 | 99 | 99 | 99 | 99 | 99 | 99 | 99 | 99 |
| SC 0.25 | 98 | 97 | 98 | 98 | 98 | 98 | 98 | 97 | 98 | 97 | 97 | GN(0,0.001) | 99 | 99 | 99 | 99 | 99 | 99 | 99 | 99 | 99 | 99 | 99 |
| SC 0. 5 | 99 | 98 | 99 | 99 | 99 | 99 | 99 | 99 | 99 | 99 | 99 | GN(0,0.01) | 99 | 99 | 99 | 99 | 98 | 99 | 99 | 99 | 99 | 99 | 99 |
| SC 0.75 | 99 | 99 | 99 | 99 | 99 | 99 | 99 | 99 | 99 | 99 | 99 | GN(0,0.06) | 97 | 98 | 97 | 97 | 96 | 97 | 97 | 97 | 97 | 98 | 98 |
| SC 1.25 | 99 | 99 | 99 | 99 | 99 | 99 | 99 | 99 | 99 | 99 | 99 | GN(0,0.1) | 96 | 97 | 96 | 96 | 94 | 96 | 95 | 96 | 96 | 97 | 97 |
| SC 1.5 | 99 | 99 | 99 | 99 | 99 | 99 | 99 | 99 | 99 | 99 | 99 | GN(0,0.3) | 93 | 95 | 92 | 93 | 92 | 93 | 92 | 92 | 92 | 94 | 94 |
| TR(10,10) | 94 | 96 | 99 | 97 | 99 | 98 | 99 | 99 | 99 | 99 | 98 | SN 0.005 | 99 | 99 | 99 | 99 | 99 | 99 | 99 | 99 | 99 | 99 | 99 |
| TR(10,20) | 96 | 97 | 99 | 98 | 99 | 98 | 99 | 99 | 99 | 99 | 98 | SN 0.001 | 99 | 99 | 99 | 99 | 99 | 99 | 99 | 99 | 99 | 99 | 99 |
| TR(20,35) | 97 | 97 | 98 | 98 | 98 | 98 | 99 | 99 | 98 | 98 | 98 | SN 0.01 | 99 | 99 | 99 | 99 | 99 | 99 | 99 | 99 | 99 | 99 | 99 |
| TR(35,45) | 96 | 97 | 98 | 98 | 98 | 98 | 98 | 99 | 98 | 98 | 97 | SN 0.1 | 97 | 99 | 99 | 98 | 99 | 98 | 98 | 98 | 98 | 99 | 99 |
| TR(50,50) | 96 | 97 | 97 | 98 | 98 | 97 | 98 | 99 | 98 | 98 | 97 | SN 0.5 | 94 | 97 | 97 | 96 | 98 | 94 | 95 | 96 | 96 | 96 | 96 |
| SE(0.2,0.2) | 97 | 98 | 99 | 99 | 99 | 97 | 99 | 98 | 98 | 99 | 98 | SP 0.005 | 99 | 99 | 99 | 99 | 99 | 99 | 99 | 99 | 99 | 99 | 99 |
| SE(1,0.2) | 96 | 97 | 98 | 98 | 98 | 97 | 97 | 96 | 97 | 98 | 98 | SP 0.001 | 99 | 99 | 99 | 99 | 99 | 99 | 99 | 99 | 99 | 99 | 99 |
| SE(0.2,1) | 96 | 98 | 98 | 97 | 96 | 96 | 96 | 97 | 96 | 97 | 97 | SP 0.01 | 99 | 99 | 99 | 99 | 99 | 99 | 99 | 99 | 99 | 99 | 99 |
| SE(0.3,0.1) | 96 | 98 | 98 | 99 | 99 | 98 | 98 | 97 | 98 | 98 | 98 | SP 0.04 | 99 | 99 | 99 | 99 | 98 | 99 | 98 | 98 | 99 | 99 | 99 |
| FL horizontal | 98 | 98 | 99 | 99 | 98 | 99 | 98 | 99 | 99 | 99 | 99 | SP 0.1 | 98 | 98 | 98 | 98 | 97 | 98 | 98 | 97 | 98 | 99 | 98 |
| FL vertical | 98 | 98 | 99 | 99 | 98 | 99 | 98 | 99 | 99 | 99 | 99 | SP 0.5 | 92 | 95 | 91 | 93 | 91 | 92 | 90 | 91 | 92 | 94 | 93 |

In this work, two other schemes also are implemented based on framework of the proposed method. In these schemes which are called DWT- BSVD and DST-BSVD, host image is decomposed using DWT and DST respectively and then watermark image is embedded in bidiagonal singular values of the sub-band from decomposition step. Same attacks are performed on these two methods like the proposed one and then PSNR, SSIM and NC are calculated for each one. In table 3 and table 4, transparency and robustness of the proposed method with the two aforementioned schemes are compared respectively. Results show that combination of DST with DWT keeps the image transparent and outperforms than in terms of robustness against various attacks.

**Table3. Transparency comparison of the proposed method with two other schemes implemented in the same framework of the DST-BSVD.**

| Image | DST-BSVD | | DWT-BSVD | | DWT-DST | |
|---|---|---|---|---|---|---|
| | SSIM | PSNR | SSIM | PSNR | SSIM | PSNR |
| F16 | 1 | 66.32 | 0.9999 | 61.6 | **1** | **69.14** |
| baboon | 1 | 66.60 | 1 | 63.23 | **1** | **73.79** |
| barbara | 1 | 66.95 | 0.9999 | 61.26 | **1** | **68.27** |
| boat | 1 | 64.98 | 0.9999 | 61.36 | **1** | **69.91** |
| couple | 0.9998 | 64.19 | 0.9996 | 59.54 | **0.9999** | **65.65** |
| elaine | 1 | 65.03 | 0.9998 | 60.94 | **1** | **67.88** |
| lenna | 1 | 65.17 | 0.9999 | 61.2 | **1** | **67.54** |
| man | 1 | 66.97 | 0.9998 | 61.32 | **1** | **73.32** |
| peppers | 1 | 66.95 | 0.9999 | 61.27 | **1** | **68.25** |
| lake | 1 | 67.58 | 1 | 62.94 | **1** | **74.62** |
| sailboat | 1 | 67.14 | 0.999 | 62.48 | **0.9999** | **74.8** |

**Table4. Comparison between the proposed method and DWT-BSVD and DWT-DST in terms of robustness.**

| Attack | DST-BSVD | | | DWT- BSVD | | | DWT-DST | | |
|---|---|---|---|---|---|---|---|---|---|
| | baboon | barbara | man | baboon | barbara | man | baboon | barbara | man |
| AF 5×5 | 0.9121 | 0.8673 | 0.8933 | 0.7825 | 0.8699 | 0.8685 | **0.974** | **0.9809** | **0.9778** |
| GP 5×5 | 0.9603 | 0.9372 | 0.9513 | 0.9827 | 0.9872 | 0.9898 | **0.9916** | **0.9926** | **0.9917** |
| MF 5×5 | 0.9345 | 0.9093 | 0.9408 | 0.8293 | 0.9332 | 0.9565 | **0.9795** | **0.9874** | **0.9852** |
| CR 30% | 0.9025 | 0.5606 | 0.8723 | 0.2528 | 0.3507 | 0.4239 | 0.7931 | 0.4489 | 0.5566 |
| CR 50% | 0.8835 | 0.8043 | 0.9667 | 0.5087 | 0.4331 | 0.4954 | **0.9224** | 0.759 | 0.8416 |
| CR 70% | 0.9532 | 0.8906 | 0.9741 | 0.6537 | 0.5513 | 0.6591 | **0.9728** | 0.8429 | 0.9422 |
| CR 90% | 0.9718 | 0.929 | 0.9636 | 0.7716 | 0.6581 | 0.7838 | 0.9698 | **0.8311** | **0.9639** |
| RO 45° | 0.8162 | 0.8191 | 0.904 | 0.9149 | 0.9897 | 0.9819 | **0.9908** | 0.9833 | 0.9665 |
| RO 70° | 0.9193 | 0.959 | 0.9762 | 0.9424 | 0.9894 | 0.9828 | **0.9877** | **0.98** | 0.9731 |
| RO -10° | 0.9462 | 0.9684 | 0.9811 | 0.9645 | 0.9919 | 0.9849 | **0.9834** | 0.9889 | **0.9869** |
| SC 0.25 | 0.9131 | 0.8735 | 0.8919 | 0.7426 | 0.8468 | 0.8308 | **0.9722** | **0.9823** | **0.9778** |
| TR(10,10) | 0.9431 | 0.9034 | 0.9492 | 0.9905 | 0.9919 | 0.9916 | 0.9688 | 0.9911 | **0.9938** |
| SE(0.3,0.1) | 0.938 | 0.9275 | 0.9626 | 0.9444 | 0.9864 | 0.9835 | **0.9804** | 0.9835 | 0.9721 |
| BL 0.2 | 0.9653 | 0.9431 | 0.9561 | 0.9928 | 0.9946 | 0.9947 | **0.9939** | 0.9938 | 0.993 |
| HE | 0.9505 | 0.9702 | 0.9659 | 0.9639 | 0.9818 | 0.9941 | **0.9694** | **0.9846** | **0.9957** |
| MB(15,45) | 0.8713 | 0.7658 | 0.8063 | 0.7573 | 0.88 | 0.8743 | 0.7615 | 0.7835 | 0.7559 |
| SH 0.1 | 0.9892 | 0.9752 | 0.9828 | 0.8131 | 0.9769 | 0.9807 | **0.9899** | **0.9961** | **0.9961** |
| SH 0.2 | 0.9901 | 0.975 | 0.9825 | 0.9572 | 0.9774 | 0.9811 | **0.9902** | **0.9962** | **0.9962** |
| SH 0.5 | 0.9912 | 0.9758 | 0.9828 | 0.9583 | 0.979 | 0.9818 | 0.9911 | **0.9964** | **0.9963** |
| SH 0.8 | 0.9913 | 0.9764 | 0.9826 | 0.9602 | 0.9797 | 0.9822 | **0.9915** | **0.9965** | **0.9964** |
| SH 1 | 0.9912 | 0.9761 | 0.9827 | 0.9619 | 0.9801 | 0.9824 | **0.9916** | **0.9966** | **0.9965** |
| JPEG Q=10 | 0.9726 | 0.9456 | 0.9558 | 0.9624 | 0.9937 | 0.9954 | **0.994** | **0.9949** | 0.9942 |
| JPEG Q=30 | 0.9685 | 0.9453 | 0.9558 | 0.9905 | 0.9942 | 0.9954 | **0.9936** | 0.9937 | 0.9935 |
| GN(0,0.06) | 0.9832 | 0.9832 | 0.988 | 0.9917 | 0.9219 | 0.9022 | 0.9854 | 0.9787 | 0.9769 |
| GN(0,0.1) | 0.9768 | 0.985 | 0.982 | 0.9689 | 0.9084 | 0.8843 | 0.9714 | 0.9626 | 0.9629 |
| GN(0,0.3) | 0.9474 | 0.9643 | 0.9514 | 0.955 | 0.8647 | 0.8367 | **0.9577** | 0.9251 | 0.9211 |
| SN 0.01 | 0.9701 | 0.9515 | 0.9713 | 0.9255 | 0.9818 | 0.9703 | **0.9956** | **0.9956** | **0.9944** |
| SN 0.1 | 0.9867 | 0.9768 | 0.9907 | 0.9946 | 0.9483 | 0.9288 | 0.9929 | **0.991** | 0.9878 |
| SN 0.5 | 0.9588 | 0.9867 | 0.9733 | 0.9828 | 0.9058 | 0.8962 | 0.9744 | 0.9726 | 0.9667 |
| SP 0.04 | 0.9804 | 0.9699 | 0.9887 | 0.9446 | 0.9532 | 0.9303 | **0.9938** | **0.9916** | **0.9895** |
| SP 0.1 | 0.9845 | 0.9713 | 0.9864 | 0.9886 | 0.9301 | 0.9091 | 0.9886 | **0.9848** | 0.9795 |
| SP 0.5 | 0.9288 | 0.9515 | 0.9519 | 0.9774 | 0.8504 | 0.8223 | 0.9523 | 0.9175 | 0.9162 |

## 4.2 Comparative analysis

In this section, our proposed DWT-DST scheme is compared with other methods in terms of imperceptibility and robustness. Table 5, compares transparency with Rawat [10], Musrrat [6], Makbol [8], Loukhaoukha [9], Mishra et.al.[12], Bhatnagar et.al.[13] and Ahmaderaghi et.al. [17].

**Table5. Transparency comparison of the proposed DST-BSVD with other studies for variety of host images that watermarked by a 256×256 gray-scale image.**

| PSNR | F16 | | baboon | | barbara | | boat | | lenna | | man | | peppers | | sailboat | |
|---|---|---|---|---|---|---|---|---|---|---|---|---|---|---|---|---|
| Rawat et.al.[10] | - | | 38.88 | | 39.05 | | - | | - | | - | | - | | - | |
| Musrrat et.al.[6] | - | | - | | - | | - | | - | | 34.87 | | - | | 33.28 | |
| Makbol et.al.[8] | - | | 55.97 | | - | | - | | 54.03 | | - | | 54.15 | | - | |
| Loukhaoukha et.al.[9] | - | | 52.37 | | - | | 54.81 | | 47.71 | | 50.18 | | 48.09 | | - | |
| Mishra et.al.[12] | - | | 50.76 | | - | | 51.49 | | 55.72 | | 51.17 | | 52.15 | | - | |
| Bhatnagar et.al.[13] | - | | - | | 35.76 | | 43.37 | | - | | - | | - | | - | |
| SSIM | PSNR | SSIM | PSNR | SSIM | PSNR | SSIM | PSNR | SSIM | PSNR | SSIM | PSNR | SSIM | PSNR | SSIM | PSNR | SSIM |
| Ahmaderaghi et.al.[17] | - | 0.999 | - | 0.999 | - | 0.999 | - | 0.999 | - | 0.999 | - | 0.999 | - | 0.999 | - | 0.999 |
| DWT-DST | **69.14** | **1** | **73.79** | **1** | **68.27** | **1** | **69.91** | **1** | **67.54** | **1** | **73.32** | **1** | **68.25** | **1** | **74.8** | **1** |

As shown in table 5, our DWT-DST scheme achieved much better imperceptibilty compared to other methods. In table 6, NC values of DWT-DST is compared with Makbol [8] and Mishra [12] approaches.

**Table6. Comparative robustness analysis by NC values between proposed DST-DWT and Makbol[8] and Mishra[12] for different attacks on sample host images that watermarked by a 256×256 gray-scale image.**

| Attack | lenna | | | baboon | | | man | |
|---|---|---|---|---|---|---|---|---|
| | proposed | [8] | [12] | Proposed | [8] | [12] | proposed | [12] |
| Pepper&salt noise(0.001) | 0.9941 | 0.994 | - | 0.9941 | 0.998 | - | 0.9941 | - |
| Pepper&salt noise(0.05) | **0.9862** | - | 0.979 | **0.9936** | - | 0.991 | **0.9884** | 0.989 |
| Gaussian noise(0.005) | **0.994** | 0.925 | - | **0.9959** | 0.732 | - | 0.9936 | - |
| Gaussian filtering(3×3) | **0.9927** | 0.987 | 0.991 | 0.9916 | 0.977 | 0.993 | **0.9918** | 0.958 |
| Speckle noise(0.4) | **0.934** | 0.704 | - | **0.9764** | 0.808 | - | 0.9681 | - |
| Rotation(45) | 0.968 | 0.983 | - | **0.9908** | 0.931 | - | 0.9665 | - |
| JPEG compression Q=40 | **0.9935** | 0.988 | - | **0.9937** | 0.915 | - | 0.9939 | - |
| JPEG compression Q=5 | 0.996 | 0.952 | 1 | 0.9909 | 0.991 | 1 | 0.9944 | 1 |
| Crop 30% | 0.9882 | - | - | 0.7931 | - | - | 0.5566 | - |
| Crop 12.5% | - | - | 0.49 | - | - | 0.699 | - | 0.724 |
| Histogram equalization | 0.98 | 0.990 | 0.994 | 0.9694 | 0.976 | 0.999 | **0.9957** | 0.987 |
| Sharpening 0.2 | **0.997** | 0.9255 | 0.995 | 0.9902 | 0.863 | 0.991 | **0.9962** | 0.9798 |

According to table 6, robustness of our DWT-DST is rather good comparing with other schemes. It is more robust against noising attacks, filtering and sharpening. In other attacks, it has close NC values, too. So it can be concluded that proposed method is almost robust against various types of attacks.

## 5 Conclusion and further work

In this study a blind image watermarking method is introduced using DWT and DST and a matrix factorization named BSVD. Multi-directional properties of DST make this scheme suitable for various images having different textures. Applying two frequency transforms, DWT and DST, sequentially on the image to decompose it, make this approach robust against different attacks. Since each of DWT and DST are resistant against some types of attacks separately, combination of them cover weakness of other transform against some attacks. So, results of tests on proposed method exhibit high transparency and acceptable robustness. Because this method is blind, it can be applicable in copyright protection.

In further research some extensions of Wavelet transform can be used in the same framework for image watermarking and analyze these transforms according to transparency, robustness and type of images. Also, other matrix factorization algorithms like QR, Lower Upper (LU) and Schur can be used in combination with Wavelet-like transforms in place of BSVD.

## 6 References

[1] Z. Shao, Y. Duan, G. Coatrieux, J. Wu, J. Meng, and H. Shu, "Combining double random phase encoding for color image watermarking in quaternion gyrator domain," *Opt. Commun.*, vol. 343, pp. 56–65, 2015.

[2] P. P. Thulasidharan and M. S. Nair, "QR code based blind digital image watermarking with attack detection code," *AEU-International J. Electron. Commun.*, vol. 69, no. 7, pp. 1074–1084, 2015.

[3] K. Ramanjaneyulu and K. Rajarajeswari, "Wavelet-based oblivious image watermarking scheme using genetic algorithm," *IET image Process.*, vol. 6, no. 4, pp. 364–373, 2012.

[4] N. M. Charkari, M. A. Z. Chahooki, and M. Radmanesh, "A Novel Approach to a High Capacity Data Hiding in Digital Images," in *Signal Processing and Information Technology, 2007 IEEE International Symposium on*, 2007, pp. 361–364.

[5] N. M. Charkari and M. A. Z. Chahooki, "A robust high capacity watermarking based on DCT and spread spectrum," in *Signal Processing and Information Technology, 2007 IEEE International Symposium on*, 2007, pp. 194–197.

[6] N. M. Makbol and B. E. Khoo, "A new robust and secure digital image watermarking scheme based on the integer wavelet transform and singular value decomposition," *Digit. Signal Process.*, vol. 33, pp. 134–147, 2014.

[7] M. Ali and C. W. Ahn, "An optimized watermarking technique based on self-adaptive DE in DWT--SVD transform domain," *Signal Processing*, vol. 94, pp. 545–556, 2014.

[8] P. Mitra and R. Gunjan, "A statistical property based image watermarking using permutation and CT-QR," in *Computer Communication and Informatics (ICCCI), 2013 International Conference on*, 2013, pp. 1–6.

[9] H. R. Kaviani, N. Karimi, and S. Samavi, "Robust watermarking in singular values of contourlet coefficients," in *Machine Vision and Image Processing (MVIP), 2011 7th Iranian*, 2011, pp. 1–5.

[10] S. R. Chalamala, K. R. Kakkirala, and R. G. B. Mallikarjuna, "Analysis of wavelet and contourlet transform based image watermarking techniques," in *Advance Computing Conference (IACC), 2014 IEEE International*, 2014, pp. 1122–1126.

[11] N. M. Makbol and B. E. Khoo, "Robust blind image watermarking scheme based on redundant discrete wavelet transform and singular value decomposition," *AEU-International J. Electron. Commun.*, vol. 67, no. 2, pp. 102–112, 2013.

[12] K. Loukhaoukha, J.-Y. Chouinard, and M. H. Taieb, "Optimal image watermarking algorithm based on LWT-SVD via multi-objective ant colony optimization," *J. Inf. Hiding Multimed. Signal Process.*, vol. 2, no. 4, pp. 303–319, 2011.

[13] N. M. Makbol and B. E. Khoo, "A new robust and secure digital image watermarking scheme based on the integer wavelet transform and singular value decomposition," *Digit. Signal Process.*, vol. 33, pp. 134–147, 2014.

[14] S. Rawat and B. Raman, "Best tree wavelet packet transform based copyright protection scheme for digital images," *Opt. Commun.*, vol. 285, no. 10, pp. 2563–2574, 2012.

[15] A. Mansouri, A. M. Aznaveh, and F. T. Azar, "SVD-based digital image watermarking using complex wavelet transform," *Sadhana*, vol. 34, no. 3, pp. 393–406, 2009.

[16] M. N. Do and M. Vetterli, "The finite ridgelet transform for image representation," *Image Process. IEEE Trans.*, vol. 12, no. 1, pp. 16–28, 2003.

[17] M. N. Do and M. Vetterli, "The contourlet transform: an efficient directional multiresolution image representation," *Image Process. IEEE Trans.*, vol. 14, no. 12, pp. 2091–2106, 2005.

[18] J.-L. Starck, E. J. Candès, and D. L. Donoho, "The curvelet transform for image denoising," *Image Process. IEEE Trans.*, vol. 11, no. 6, pp. 670–684, 2002.

[19] G. Easley, D. Labate, and W.-Q. Lim, "Sparse directional image representations using the discrete shearlet transform," *Appl. Comput. Harmon. Anal.*, vol. 25, no. 1, pp. 25–46, 2008.

[20] M. N. Do and M. Vetterli, "Image denoising using orthonormal finite ridgelet transform," in *International Symposium on Optical Science and Technology*, 2000, pp. 831–842.

[21] R. Eslami and H. Radha, "The contourlet transform for image denoising using cycle spinning," in *Signals, Systems and Computers, 2004. Conference Record of the Thirty-Seventh Asilomar Conference on*, 2003, vol. 2, pp. 1982–1986.

[22] G. R. Easley, D. Labate, and F. Colonna, "Shearlet-based total variation diffusion for denoising," *Image Process. IEEE Trans.*, vol. 18, no. 2, pp. 260–268, 2009.

[23] M. N. Do and M. Vetterli, "Orthonormal finite ridgelet transform for image compression," in *Image Processing, 2000. Proceedings. 2000 International Conference on*, 2000, vol. 2, pp. 367–370.

[24] Y. Li, Q. Yang, and R. Jiao, "Image compression scheme based on curvelet transform and support vector machine," *Expert Syst. Appl.*, vol. 37, no. 4, pp. 3063–3069, 2010.

[25] S. AlZubi, N. Islam, and M. Abbod, "Multiresolution analysis using wavelet, ridgelet, and curvelet transforms for medical image segmentation," *J. Biomed. Imaging*, vol. 2011, p. 4, 2011.

[26] Y. Sha, L. Cong, Q. Sun, and L. Jiao, "Unsupervised image segmentation using contourlet domain hidden markov trees model," in *Image Analysis and Recognition*, Springer, 2005, pp. 32–39.

[27] A. Mosleh, F. Zargari, and R. Azizi, “Texture image retrieval using contourlet transform,” in *Signals, Circuits and Systems, 2009. ISSCS 2009. International Symposium on*, 2009, pp. 1–4.

[28] I. J. Sumana, M. M. Islam, D. Zhang, and G. Lu, “Content based image retrieval using curvelet transform,” in *Multimedia Signal Processing, 2008 IEEE 10th Workshop on*, 2008, pp. 11–16.

[29] A. B. Gonde, R. P. Maheshwari, and R. Balasubramanian, “Multiscale ridgelet transform for content based image retrieval,” in *Advance Computing Conference (IACC), 2010 IEEE 2nd International*, 2010, pp. 139–144.

[30] G. Y. Chen, T. D. Bui, and A. Krzy\.zak, “Rotation invariant feature extraction using Ridgelet and Fourier transforms,” *Pattern Anal. Appl.*, vol. 9, no. 1, pp. 83–93, 2006.

[31] Z. Zhang, S. Ma, and X. Han, “Multiscale feature extraction of finger-vein patterns based on curvelets and local interconnection structure neural network,” in *Pattern Recognition, 2006. ICPR 2006. 18th International Conference on*, 2006, vol. 4, pp. 145–148.

[32] A. Bouzidi and N. Baaziz, “Contourlet domain feature extraction for image content authentication,” in *Multimedia Signal Processing, 2006 IEEE 8th Workshop on*, 2006, pp. 202–206.

[33] S. Zhou, J. Shi, J. Zhu, Y. Cai, and R. Wang, “Shearlet-based texture feature extraction for classification of breast tumor in ultrasound image,” *Biomed. Signal Process. Control*, vol. 8, no. 6, pp. 688–696, 2013.

[34] E. J. Leavline, S. Sutha, and D. A. A. G. Singh, “ON THE SUITABILITY OF MULTISCALE IMAGE REPRESENTATION SCHEMES AS APPLIED TO NOISE REMOVAL.”

[35] B. Ahmederahgi, F. Kurugollu, P. Milligan, and A. Bouridane, “Spread spectrum image watermarking based on the discrete shearlet transform,” in *Visual Information Processing (EUVIP), 2013 4th European Workshop on*, 2013, pp. 178–183.

[36] S. Yi, D. Labate, G. R. Easley, and H. Krim, “A shearlet approach to edge analysis and detection,” *Image Process. IEEE Trans.*, vol. 18, no. 5, pp. 929–941, 2009.

[37] J. Fadili and J.-L. Starck, “Curvelets and ridgelets,” in *Computational Complexity*, Springer, 2012, pp. 754–773.

[38] B. Ahmaderaghi, J. Martinez Del Rincon, F. Kurugollu, and A. Bouridane, “Perceptual watermarking for Discrete Shearlet transform,” in *Visual Information Processing (EUVIP), 2014 5th European Workshop on*, 2014, pp. 1–6.

[39] M. Amini and H. Sadreazami, “Binary image watermarking in ridgelet domain,” in *Signal Processing (ICSP), 2010 IEEE 10th International Conference on*, 2010, pp. 1813–1816.

[40] N. K. Kalantari, S. M. Ahadi, and M. Vafadust, “A robust image watermarking in the ridgelet domain using universally optimum decoder,” *Circuits Syst. Video Technol. IEEE Trans.*, vol. 20, no. 3, pp. 396–406, 2010.

[41] P. Mangaiyarkarasi and S. Arulselvi, “A new digital image watermarking based on Finite Ridgelet Transform and extraction using ICA,” in *Emerging Trends in Electrical and Computer Technology (ICETECT), 2011 International Conference on*, 2011, pp. 837–841.

[42] H. Song and J. Gu, "Curvelet based adaptive watermarking for images," in *Computer Science and Network Technology (ICCSNT), 2012 2nd International Conference on*, 2012, pp. 1101–1105.

[43] Q. Jian-ying, S. Yuan-tong, N. Yan-bin, and F. Lei, "Digital watermarking algorithms based on curvelet transform," in *2011 International Conference on Electric Information and Control Engineering*, 2011, pp. 4150–4153.

[44] N. M. Kumar, T. Manikandan, and V. Sapthagirivasan, "Non blind image watermarking based on similarity in contourlet domain," in *Recent Trends in Information Technology (ICRTIT), 2011 International Conference on*, 2011, pp. 1277–1282.

[45] H. Sadreazami and M. Amini, "Highly robust image watermarking in contourlet domain using singular value decomposition," in *Signal Processing (ICSP), 2012 IEEE 11th International Conference on*, 2012, vol. 1, pp. 628–631.

[46] S. Zhu and J. Liu, "Adaptive image watermarking scheme in contourlet transform using singular value decomposition," in *Advanced Communication Technology, 2009. ICACT 2009. 11th International Conference on*, 2009, vol. 2, pp. 1216–1219.

[47] S. Majumder, M. Saikia, T. S. Das, and S. K. Sarkar, "Hybrid image watermarking scheme using SVD and PDFB based contourlet transform," in *Computer and Communication Technology (ICCCT), 2011 2nd International Conference on*, 2011, pp. 130–134.

[48] H. Yingkun, Z. Chunxia, L. Mingxia, and Y. Deyun, "The nonsubsampled contourlet-wavelet hybrid transform: Design and application to image watermarking," in *Computer Science and Software Engineering, 2008 International Conference on*, 2008, vol. 3, pp. 627–630.

[49] C.-K. Chan and L.-M. Cheng, "Hiding data in images by simple LSB substitution," *Pattern Recognit.*, vol. 37, no. 3, pp. 469–474, 2004.

[50] H. S. Prasantha, H. L. Shashidhara, and K. N. Balasubramanya Murthy, "Image compression using SVD," in *Conference on Computational Intelligence and Multimedia Applications, 2007. International Conference on*, 2007, vol. 3, pp. 143–145.

[51] P. K. Sadasivan and D. N. Dutt, "SVD based technique for noise reduction in electroencephalographic signals," *Signal Processing*, vol. 55, no. 2, pp. 179–189, 1996.

[52] R. Liu and T. Tan, "An SVD-based watermarking scheme for protecting rightful ownership," *Multimedia, IEEE Trans.*, vol. 4, no. 1, pp. 121–128, 2002.

[53] G. Bhatnagar and B. Raman, "Robust reference-watermarking scheme using wavelet packet transform and bidiagonal-singular value decomposition," *Int. J. Image Graph.*, vol. 9, no. 03, pp. 449–477, 2009.

[54] C.-C. Lai and C.-C. Tsai, "Digital image watermarking using discrete wavelet transform and singular value decomposition," *Instrum. Meas. IEEE Trans.*, vol. 59, no. 11, pp. 3060–3063, 2010.

[55] S. Som, S. Palit, K. Dey, D. Sarkar, J. Sarkar, and K. Sarkar, "A DWT-based Digital Watermarking Scheme for Image Tamper Detection, Localization, and Restoration," in *Applied Computation and Security Systems*, Springer, 2015, pp. 17–37.

[56] X. Gibert, V. M. Patel, D. Labate, and R. Chellappa, “Discrete shearlet transform on GPU with applications in anomaly detection and denoising,” *EURASIP J. Adv. Signal Process.*, vol. 2014, no. 1, pp. 1–14, 2014.

[57] K. Xu, S. Liu, and Y. Ai, “Application of Shearlet transform to classification of surface defects for metals,” *Image Vis. Comput.*, vol. 35, pp. 23–30, 2015.

[58] X. Sun, X. Wu, and Y. Wei, “Face Recognition Based on Shearlet Transform and Fast ICA,” in *Instrumentation and Measurement, Computer, Communication and Control (IMCCC), 2014 Fourth International Conference on*, 2014, pp. 832–835.

[59] ShearLab 3D (version 1.1)[Computer software], Retrieved from http://www.shearlet.org/